\documentstyle[preprint,aps]{revtex}

\begin{document}
\draft
\preprint{FNUSAL - 3/96}

\title{Mass and width of the $d'$ resonance in nuclei}

\author{A. Valcarce, H. Garcilazo\thanks{
On leave from Escuela Superior de F\' \i sica y Matem\'aticas,
Instituto Polit\'ecnico Nacional, Edificio 9,
07738 M\'exico D.F., Mexico},
and F. Fern\'andez}

\address{ Grupo de F\' \i sica Nuclear \\
Universidad de Salamanca, E-37008 Salamanca, Spain}

\date{\today}

\maketitle

\begin{abstract}
We calculated the mass and width of the $d'$ resonance
inside nuclei within a nucleon-$\Delta$ model
by including the self-energy of the $\Delta$
in the $N\Delta$ propagator. We found that
in the nuclear medium the width
of the $d'$ is increased by one order of magnitude
while its mass changes only by a few MeV.
This broadening of the width of the $d'$
resonance embedded in nuclei is consistent with
the experimental observations so that
the $d'$ can be understood as
a $N\Delta$ resonance. Thus,
given the freedom between either
isospin 0 or isospin 2 for the $d'$,
our results
give weigth to the isospin-2 assignment.
\end{abstract}


\section{Introduction}

The possible existence of a resonance with quantum numbers
$J^P=0^-$ and isospin even has been inferred from
the differential cross section of the
double charge exchange (DCX) reaction \cite{BIL1} in
nuclei ranging from
$^{14}$C to $^{48}$Ca.
The authors of ref. \cite{BIL1}
chose the isospin-0 assignment for the resonance showing up
in the DCX reaction based on the predictions of a QCD
string model \cite{MUL}.
It was soon pointed out \cite{GAR1}
that if the $d'$ resonance decays into a pion and two nucleons,
then it should be a solution of the equations describing the dynamics
of the $\pi NN$ system. That this is not the case for
isospin 0 was already
known before \cite{GAR2}, since the dominant interaction of the
$\pi NN$ system in the isospin-0 sector is the pole part of the
$\pi N$ $P_{11}$ channel.
This interaction is forbidden by the Pauli
principle in the case of the $J^P=0^-$ channel and as a consequence
the $0^-$ channel is very weak (and repulsive \cite{GAR2})
so that no resonance
can arise there. The situation of
the isospin-2 sector
is quite different. In this case, the dominant interaction is the
$\pi N$ $P_{33}$ channel (the $\Delta$ resonance), and all
existing conventional calculations \cite{GAR2,VGAR,GAL,UED,KAL}
find that the $0^-$ channel is very attractive so that a resonance
could exist in that channel.
This led the authors of
ref. \cite{GAR1} to propose the isospin-2 assignment of the
$d'$. Other calculations based on completely different dynamical
models like the Skyrme model \cite{SCO} or six-quark bags
\cite{GLO} have arrived to similar conclusions.

The known experimental facts of the $d'$ resonance are
the following. From the analysis
of the DCX data performed in ref. \cite{BIL1},
its free width and its mass and
width in the nuclear medium were extracted.
They found for these quantities
$\Gamma_{\rm Free}$=0.51 MeV, $M$=2065 MeV,  and
$\Gamma_{\rm Medium}$= 5 MeV (they
actually found it to be between 3 and 7 MeV
and took the average value)
for nuclei ranging from $^{14}$C to $^{48}$Ca.
No value was given for the free mass of the $d'$.
They also observed that the mass of the $d'$ in the
medium is almost independent of the nucleus.
We should stress that these values were
not obtained from a given theoretical model
of the $d'$ but were simply extracted
from the experimental data.

In order to see if the main features of the $d'$
can be predicted by a theory involving pions, nucleons,
and $\Delta$'s, we have launched a study of the isospin-2
sector using as basic framework the nucleon-$\Delta$
interaction derived from the chiral quark cluster model
\cite{JOUR,PED1,PED2}. In ref. \cite{VGAR} we found that
the $0^-$ channel is the most attractive channel of the
isospin-2 sector. However, since in that calculation we
neglected the non-local terms of the $N\Delta$ interaction
as well as the contribution of the $\Delta \Delta$ channels,
we found that the existing attraction was not enough to give rise
to a resonance. In order to simulate the lacking attraction
we varied the mass of the sigma meson, that is responsible
for the intermediate range attraction, until a
resonance of mass 2064.4 MeV was produced.
We found that the free width
of the resonance is 0.6 MeV, in agreement
with the value extracted in ref. \cite{BIL1}. The problem of the
lack of attraction of our model has been partially solved in
ref. \cite{ESTE} by including the non-local terms of the $N\Delta$
interaction. This improved model is able to produce a resonance
although with a mass 80 MeV higher than the
experimentally observed one.
In order to get the mass of the
resonance down to its correct value it
probably will be necessary to include
the contribution of the $\Delta \Delta$ channels. As is well
known \cite{LOMON}, the inclusion of coupled channels
with a threshold above the energy in consideration leads always to
additional attraction. The main result obtained
in these works was
that the width of the $d'$ is strongly correlated with its mass
such that if the mass of the $d'$ approaches the $\pi NN$ threshold
then the width drops very fast to very small values.

Our aim in this paper is to try to understand
the effects of the nuclear
medium in the properties of the $d'$ resonance. In Sec. II
we will briefly review our formalism and its extension to
the case when the $d'$ is inside a nucleus. Section III will be
devoted to the concept of the self-energy of the $\Delta$ due
to the presence of a nuclear medium. We present our results in Sec. IV,
and we conclude with their discussion in Sec. V.

\section{Formalism}

Since there is considerable theoretical evidence that a
resonance with isospin 2 exists in the $J^P=0^-$ channel
\cite{GAR2,GAL,UED,KAL,SCO,GLO}, we have calculated the
mass and width of the $d'$ resonance in refs. \cite{VGAR,ESTE}
by solving the Lippmann-Schwinger equation for the $N\Delta$
system,

\begin{equation}
T_{ij}(\sqrt{S} ; {\vec q} \, ',{\vec q}_0) \, = \,
V_{ij}({\vec q} \, ',{\vec q}_0) \, + \, \sum_k \int d {\vec q} \, \,
V_{ik}({\vec q} \, ' ,{\vec q} \, ) G_0(\sqrt{S},q)
T_{kj}(\sqrt{S} ; {\vec q},{\vec q}_0) \, ,
\end{equation}

\noindent
where $V_{ik}({\vec q}\, ',{\vec q} \, )$ are the (in general non-local)
$N\Delta$ interactions predicted by the chiral-quark-cluster model
\cite{JOUR,PED1,PED2}. The propagator of the $N\Delta$ intermediate
state is \cite{VGAR}

\begin{equation}
G_{\Delta}(\sqrt{S},q)={2 M_\Delta \over
s-M_\Delta^2 + i M_\Delta \Gamma_\Delta(q)} \, ,
\end{equation}

\noindent
where $\sqrt{S}$ is the invariant mass of the system,
$q$ is the magnitude of the $N\Delta$
relative momentum and $s$ is the invariant mass
squared of the $\pi N$ subsystem given by

\begin{equation}
s=S+M_N^2-2  \sqrt{S(M_N^2+q^2)} \, .
\end{equation}

\noindent
The width $\Gamma_\Delta(q)$ was taken as

\begin{equation}
\Gamma_\Delta(q)={2 \over 3}\,0.35\, p_0^3 {\sqrt{M_N^2+q^2}\over
m_\pi^2\sqrt{s}} \, ,
\end{equation}

\noindent
where

\begin{equation}
p_0=\left([s-(M_N+m_\pi)^2][s-(M_N-m_\pi)^2] \over 4s \right)^{1/2} \, ,
\end{equation}

\noindent
is the magnitude of the pion-nucleon relative momentum.

From the solution of these equations in the case when they
give rise to a resonance, we obtained the properties of the
$d'$ (mass and width) in free space. As already mentioned,
we found in ref. \cite{VGAR} that the $J^P=0^-$ channel
is the most attractive one in the isospin-2 sector.
In our model this attraction originates from several facts.
First of all, the quark Pauli blocking giving
rise to a very strong repulsive barrier at short distances
is not present in this channel \cite{PED1,PED2}. Secondly, the
dominant terms of the interaction at intermediate
range, i.e., the sigma exchange
and the tensor part of the one-pion exchange are both
attractive. Finally, the contribution from
the central part of the one-pion exchange
and the gluons is
rather weak, the last one being repulsive but
restricted to very small distances.
Notice that the short-range
behavior has almost no effect in this
channel since the nucleon and the $\Delta$ are in a relative P-wave
state.

We will now extend our
formalism in order to study how the properties of the $d'$ (mass
and width) are modified when it is embedded in a nuclear medium.
The $d'$ resonance behaves very differently inside a nucleus
than in free space. In particular, in free space the $d'$
cannot decay into two nucleons, while inside a nucleus
such process is possible as show in Fig. 1 for a
typical diagram. The main physical concept necessary
in order to describe theoretically the influence of the
medium in a $N\Delta$ system is the self-energy
of the $\Delta$ in the medium \cite{BROWN,WEISE,NIEVES,POLLS}.
This quantity has been calculated for the case of infinite
nuclear matter \cite{BROWN,WEISE,NIEVES} as well as for the case
of finite nuclei \cite{POLLS}.

In the case of a single $\Delta$ propagating through a
nuclear medium, the propagator of the $\Delta$ is given by

\begin{equation}
G_{\Delta}(E,\omega)={1 \over
E-M_E(\omega)+{i \over 2} \Gamma_E(\omega)} \, ,
\end{equation}

\noindent
where $M_E(\omega)$ and ${1 \over 2} \Gamma_E(\omega)$ are the
effective mass and one half of the
width of the $\Delta$ in the medium,

\begin{eqnarray}
M_E(\omega) & = & M_{\Delta} \, + \,
Re \Sigma_{\Delta}(\omega) \, ,   \\
{1 \over 2} \Gamma_E(\omega) & = & {1 \over 2} \Gamma_{\Delta} \, - \,
Im \Sigma_{\Delta}(\omega) \, ,
\end{eqnarray}

\noindent
with $M_\Delta$ and $\Gamma_\Delta$ being the free mass and width
of the $\Delta$ respectively, and $Re \Sigma_{\Delta}(\omega)$
and $Im \Sigma_{\Delta}(\omega)$ the real and imaginary parts
of the self-energy of the $\Delta$ in the medium which are functions
of the pion energy $\omega$ (see next section).

Thus, in order to study the effects
of the nuclear medium in the
properties of the $d'$ we solved the Lippmann-Schwinger
equation (1) including
the self-energy of the
$\Delta$ in the $N\Delta$ propagator (2),
i.e., we made in that equation the replacements

\begin{eqnarray}
M_\Delta & \to & M_\Delta + Re \Sigma_\Delta(\omega) \, , \\
{1 \over 2} \Gamma_\Delta (q) & \to & {1 \over 2} \Gamma_\Delta (q) -
Im \Sigma_\Delta(\omega) \, ,
\end{eqnarray}

\noindent
with

\begin{equation}
\omega \, = \, \sqrt{p_0^2 + m_\pi^2} \, ,
\end{equation}

\noindent
and $p_0$ given by eq. (5).

\section{The self-energy of the $\Delta$}

The self-energy of the $\Delta$ has been calculated in infinite nuclear
matter \cite{BROWN,WEISE,NIEVES} and more recently
also for the case of finite nuclei \cite{POLLS}. The most
important quantity, with respect to our particular problem, is
the imaginary part of the self-energy $Im \Sigma_\Delta(\omega)$.
This quantity has been calculated for finite nuclei by Hjorth-Jensen,
M\"uther, and Polls \cite{POLLS} within the framework of perturbative
many-body theory using a basis of single-particle states appropriate
for both bound hole states and for particle states in the continuum.

We show in Fig. 2 the values of $|Im \Sigma_\Delta(\omega)|$
for $0 \leq \omega
\leq 200$ MeV for the three nuclei
$^{16}$O, $^{40}$Ca, and $^{100}$Sn
obtained in ref. \cite{POLLS}
together with
the result of Nieves {\it et al.} \cite{NIEVES} for infinite
nuclear matter. For comparison, we show in the same figure
$\Gamma_\Delta/2$, one half
of the free width of the $\Delta$
given by eq. (4).
The most striking feature that is observed in Fig. 2 is that
while $\Gamma_\Delta$
starts at $\omega \approx$140 MeV
(the mass of the pion), the imaginary part of the self-energies
of infinite nuclear matter and finite nuclei
starts already at $\omega =$0. This behavior simply reflects
the fact that the pion can be scattered by a nucleon only if
$\omega \ge m_\pi$ (the physical region for scattering),
while the process of absorption of a pion by a nucleus can
occur if the pion is at rest ($\omega = m_\pi$) or even if
the pion energy is smaller than its rest mass, as long as
$\omega \ge$0.
We have also drawn in Fig. 2 the maximum value of
$\omega$ that enters in eqs. (1)-(5) and (9)-(11)

\begin{equation}
\omega_{\rm{Max}} \, = \, {{(M_{d'}-M_N)^2 - M_N^2 + m_\pi^2}
\over {2 (M_{d'} - M_N)}} \, .
\end{equation}

\noindent
This
$\omega_{\rm{Max}}$ is obtained by taking
$q=$ 0 and $\sqrt{S} = M_{d'}$
in eqs. (2)-(5) and (11).
It is easy to see from these equations
that as $q$ increases $\omega$ decreases,
such that it goes from $\omega_{\rm{Max}}$ down to $- \infty$.
Also, if the mass of the $d'$ resonance decreases
the value of $\omega_{\rm{Max}}$ moves to the left
in Fig. 2.
In particular, if the mass of the resonance approaches the
$\pi NN$ threshold ($M_{d'} \to 2M_N + m_\pi$) then
$\omega_{\rm{Max}} \to m_\pi$.

The last feature that is worth noticing in Fig. 2
is that for finite nuclei $|Im \Sigma_\Delta|$ increases
with the size of the nucleus, i.e., it is largest for
$^{100}$Sn and smallest for $^{16}$O.

The real part of the self-energy
$Re \Sigma_\Delta(\omega)$ has been found to be constant and
equal to $-$53$\rho/\rho_0$ MeV for $\omega < 190$ MeV by
Nieves {\it et al.} \cite{NIEVES} in the case
of infinite nuclear matter.
This value of $Re \Sigma_\Delta \approx$ $-$50 MeV for infinite
nuclear matter has been known already for quite a long time
\cite{BROWN}.
Since in our calculation
of the $d'$ we have always $\omega < 190$ MeV,
and the calculations of ref. \cite{POLLS} correspond
to $\rho/\rho_0 = 0.75$,
we used $Re \Sigma_\Delta(\omega)= -$40 MeV.

\section{Results}

Before presenting our results it is worthwhile to discuss
the general properties of resonances predicted
by our $N \Delta$ model. Of particular importance is the
behavior of the width of a resonance when its mass is near
the $\pi NN$ threshold as it is the case of the $d'$.
Let us consider first the case when the $d'$ is in free space,
i.e., it appears as a solution of eqs. (1)-(5). Since the mass
of the $d'$ (2065 MeV) is below the $N\Delta$ threshold (2172 MeV),
then if the $\Delta$ were a stable particle the $d'$ would be
simply a bound state of the $N\Delta$ system. Thus, if we make
$\Gamma_\Delta (q) =$0 in eq. (2), then eq. (1) has a
bound-state solution, i.e., a pole, at $\sqrt{S} \approx M_{d'}$.
If we now put back the width of the $\Delta$ given by eqs. (4)
and (5) the pole moves from the real axis into the complex plane,
i.e., to the position $\sqrt{S} = M_{d'}-{i \over 2} \Gamma_{d'}$.
Thus, the $d'$ acquires its width as a direct consequence of the
width of the $\Delta$.

We have shown in Fig. 2 the width of the $\Delta$ as a function of
$\omega$. Also, in the discussion of the previous section we showed
that $\omega$ varies from $\omega_{\rm{Max}}$ down to $- \infty$, so
that only that part of $\Gamma_\Delta$ lying between
$\omega = m_\pi$ and $\omega = \omega_{\rm{Max}}$ contributes
in the integral equation (1). However, as shown by eq. (12)
$\omega_{\rm{Max}} \to m_\pi$ when $M_{d'} \to 2M_N + m_\pi$,
which means that as $M_{d'}$ gets closer to the $\pi NN$
threshold the contribution of $\Gamma_\Delta$ to eq. (1)
becomes less and less important and consequently the
width of the $d'$ tends to zero in the
limit $M_{d'} \to 2M_N + m_\pi$
(notice that $M_{d'}$ is less than 50 MeV
above the $\pi NN$ threshold). In addition,
for $\omega$ near $\omega_{\rm{Max}}$ (where
$\Gamma_\Delta$ is large) $q \to 0$ so that
here $\Gamma_\Delta$ is suppressed by a factor
$q^3$, $q^2$ coming from the volume element
in eq. (1) and $q$ coming from the interaction
(which is a P wave).
That explains
why the width of the $d'$ is so small in free space
($\sim$ 0.5 MeV).

If we now consider the case when the $d'$ is embedded in
a nuclear medium we have to make the replacements
$M_\Delta \to M_\Delta + Re \Sigma_\Delta$ and
${1 \over 2} \Gamma_\Delta \to {1 \over 2} \Gamma_\Delta -
Im \Sigma_\Delta$ in eqs. (1)-(5). Therefore, by looking
at Fig. 2 one expects that the effect of including the self-energy
of the $\Delta$ will be an increase in the width of the $d'$
by about one order of magnitude, since the area under the
curves for $|Im \Sigma_\Delta|$ is roughly one order of
magnitude larger than for ${1 \over 2} \Gamma_\Delta$
for $\omega$ between 0 and $\omega_{\rm{Max}}$.

We show in Table I the results for the
mass and width of the $d'$ resonance when it is in free space,
inside $^{16}$O, $^{40}$Ca, or $^{100}$Sn, and inside infinite
nuclear matter using our standard model of ref. \cite{VGAR}.
As one can see, the effect on the mass of the $d'$ is an
increase of 3-4 MeV. As expected from the previous discussion,
the most important effects are seen on the
width, which changes by more than one order of magnitude.
Also, as expected from Fig. 2, the broadening increases
with the size of the nucleus, being
maximum for the case of infinite nuclear matter. If we neglect
the real part of the self-energy, then the mass of the $d'$
returns to the free mass except by a few tenths of
MeV, while there is almost no change in any of
the widths of the $d'$ given in Table I.

Since by including the self-energy of the $\Delta$
the mass of
the $d'$ increases by 3-4 MeV,
in order to account for this mass shift
we have readjusted the mass of
the sigma meson from 234 MeV to 231.5 MeV in our standard model.
These results are shown
in Table II. The mass of the $d'$ in the medium is now in agreement
with the experimentally extracted value. Notice that as observed
in ref. \cite{BIL1} the mass of the $d'$ in the medium is almost
independent of the nucleus in which the $d'$ is embedded. However,
the mass and the width of the $d'$ in free space are now 2061.1 MeV
and 0.47 MeV, respectively. We should point out that in ref.
\cite{BIL1} they extracted the mass of the $d'$ embedded in
nuclei and both the free width and the width in the medium, but not
the free mass. Thus, the value 2061.1 MeV is
not in contradiction with the value of the mass given
in ref. \cite{BIL1}.
The free width is in very good agreement
with the value 0.51 MeV obtained in ref. \cite{BIL1} while the
widths in the medium for the nuclei $^{16}$O and $^{40}$Ca are
comparable to those extracted in ref. \cite{BIL1} although somewhat
higher in the case of $^{40}$Ca (they obtained
between 3 and 7 MeV).

In order to understand the overestimation of the width of the
$d'$ in $^{40}$Ca shown in Table II one should be aware of the
following. The self-energies of ref. \cite{POLLS}
are $r$ dependent,
where $r$ is the radial coordinate. They gave
the values of $Im \Sigma_\Delta(\omega)$
shown in Fig. 2 which correspond to a 'typical' radius
of 1.5 fm. While this value of $r$ is perhaps appropriate
for light nuclei (carbon, oxigen), it is not for medium or
heavy nuclei. In particular, in the case of $^{40}$Ca the wave
functions of the nucleons of the $1d_{3/2}$ and $1f_{7/2}$
shells, which are the relevant ones for the DCX reaction, peak
at around 3 fm. Therefore, if one looks at the results shown in
Fig. 2 of ref. \cite{POLLS}, one sees
that $Im \Sigma_\Delta(\omega)$ should be roughly a factor of
two smaller for $r$ around 3 fm. If we repeat the calculation
given in Table II for $^{40}$Ca with
$Im \Sigma_\Delta(\omega)$ divided by two, we get instead
of 10.30 MeV, a value of 5.43 MeV for the width of the $d'$.
This would be in good agreement with the observations of
ref. \cite{BIL1}. Using a similar argument the
shift of the mass of the $d'$ due to the
medium (see Tables I and II) would be reduced.
Since $Re \Sigma_\Delta(\omega)$ is a linear function of the
density \cite{NIEVES} and for $r \sim$ 3 fm
the nuclear density has decreased
by more than a factor two, repeating again
the calculation of Table II leads us to an estimate of
the shift of the mass of
the $d'$ due to the medium, not being  more than 2 MeV.

\section{Discussion}

Thus, as we have just shown, our model is able to give
a consistent description of the properties of the $d'$
obtained in the analysis of ref. \cite{BIL1} just by
assuming that the $d'$ is a $N\Delta$ resonance of isospin 2,
i.e., the isospin-2 assignment of the $d'$ is consistent
with everything we know about pions, nucleons, and $\Delta$'s
both when they are in free space or when they are embedded
in nuclear matter. This strongly enhances the case for the
isospin-2 assignment of the $d'$.

Finally, since in our opinion there is enough theoretical
evidence indicating a resonance with isospin 2, we like
to suggest a new experimental search of it. The reaction
$\pi^- \, d \to \pi^+ \, X$, where $X=\pi^- nn$, is best
suited for this purpose since $X$ necessarily has isospin 2.
This reaction was used in the past \cite{PIA,LICH,KAML}
to search for a possible $\pi^- nn$ bound state. In that
case, they were looking mainly at the region near the
$\pi NN$ threshold, although data were taken also for
energies above this threshold. However, in the
region of about 50 MeV above the $\pi NN$ threshold (where
the $d'$ is) they used a very large energy step ($\sim$ 15 MeV)
and a resolution of about 3 MeV so that they could not possibly
see the $d'$. In order to detect the $d'$ a resolution better
than 0.5 MeV would be needed and they should look in the mass
region between 2065 MeV and at most a couple of MeV below
this value.

\acknowledgements
We thank to H. M\"uther for useful comments.
This work has been partially funded by Direcci\'on
General de Investigaci\'on Cient\'{\i}fica y T\'ecnica
(DGICYT) under the Contract No. PB94-0080-C02-02 and by
COFAA-IPN (M\'exico).
H.G. thanks the Direcci\'on General de Investigaci\'on
Cient\' \i fica y T\'ecnica of the Spanish Ministery
for Science and Education for a sabbatical stay
(Ref.: SAB95-0338).

\begin{figure}
\caption{ Diagramatic representation of a process that gives
origin to the broadening of the $d'$ resonance in nuclei.}
\label{fig1}
\end{figure}

\begin{figure}
\caption{Imaginary part of the self-energies
of the $\Delta$ for the three nuclei
$^{16}$O, $^{40}$Ca and $^{100}$Sn \protect{\cite{POLLS}} as
well as for infinite nuclear matter \protect{\cite{NIEVES}}.
Also shown in the figure is the free width of the $\Delta$
given by eq. (4) divided by 2.}
\label{fig2}
\end{figure}

\begin{table}
\caption{ Mass and width of the $d'$ resonance in free space,
finite nuclear matter, and infinite nuclear matter
for $m_\sigma$=234.0 MeV.}
\label{width}

\begin{tabular}{ccccc}
 & Case       & $M_{d'}$ (MeV)  &  $\Gamma_{d'}$ (MeV) & \\
\tableline
 & Free space                    &  2064.4      &  0.62          & \\
 & $^{16}$O                      &  2067.8      &  6.18          & \\
 & $^{40}$Ca                     &  2068.0      &  10.98         & \\
 & $^{100}$Sn                    &  2068.5      &  16.88         & \\
 & Infinite nuclear matter       &  2068.9      &  23.96         & \\
\end{tabular}
\end{table}

\begin{table}
\caption{ Mass and width of the $d'$ resonance in free space,
finite nuclear matter, and infinite nuclear matter for
$m_\sigma$=231.5 MeV.}
\label{width2}

\begin{tabular}{ccccc}
 & Case       & $M_{d'}$ (MeV)  &  $\Gamma_{d'}$ (MeV) & \\
\tableline
 & Free space                    &  2061.1      &  0.47          & \\
 & $^{16}$O                      &  2064.6      &  5.68          & \\
 & $^{40}$Ca                     &  2064.8      &  10.30         & \\
 & $^{100}$Sn                    &  2065.2      &  15.83         & \\
 & Infinite nuclear matter       &  2065.6      &  23.06         & \\
\end{tabular}
\end{table}

\end{document}